\documentclass[a4paper, 10pt, twocolumn]{article}
\usepackage{graphicx}
\newcommand{\beq}{\begin{equation}}
\newcommand{\eeq}{\end{equation}}
\newcommand{\beqa}{\begin{eqnarray}}
\newcommand{\eeqa}{\end{eqnarray}}

\def\opone{\leavevmode\hbox{\small1\normalsize\kern-.33em1}}

\title{Quantum nonlocality: How does Nature perform the trick?\cite{Bellprize}}

\author{Nicolas Gisin \\
\it \small   Group of Applied Physics, University of Geneva, 1211 Geneva 4,    Switzerland}

\date{\small \today}

\begin{document}
\maketitle

Since our early childhood we know in our bones that in order to interact with an object we have either to go to it or to throw something at it. Yet, contrary to all our daily experience, Nature is nonlocal: there are spatially separated systems that exhibit nonlocal correlations. In recent years this led to new experiments, deeper understanding of the tension between quantum physics and relativity and to proposals for disruptive technologies.

Consider two spatially separated quantum systems, one controlled by Alice, the other by Bob, in a pure state $\psi$. Alice and Bob may perform some measurements $x$ and $y$ on their systems and collect the results $a$ and $b$, respectively. This situation is described by a conditional probability distribution $p_\psi (a,b|x,y)$. In general this correlation doesn't factorize: $p_\psi (a,b|x,y)\neq p_\psi (a|x)\cdot p_\psi (b|y)$, i.e. the two systems are correlated. At first, this is no surprise, correlations are everywhere. For example, consider two cups of the same color, either both red or both green, one in Alice's and one in Bob's hands. If they looks at the color of their cups, Alice and Bob's results are correlated. In this example the origin of the correlation is obvious, Alice and Bob had only partial information: they knew that both have the same color, but they ignored which color. This differs deeply from the quantum situation, as quantum theory claims that a pure state provides a complete description of the two systems. This led EPR\cite{EPR} to believe that quantum theory is incomplete in the same sense as the description "of the same color" provides only an incomplete description of the color state of the cups.

Let us now consider the situation described by any possible future theory. Define $\lambda$ as the state that this future theory ascribes to the two spatially separated systems and assume that: $p_\lambda (a,b|x,y)= p_\lambda (a|x)\cdot p_\lambda (b|y)$. A priori it seems hard to make any prediction from this assumption since we do not know this future theory. But John Bell noticed that the experiments we perform today necessarily correspond to a statistical mixture of the more refined states of this future theory: $p_\psi(a,b|x,y) = \int d\lambda \rho(\lambda) p_\lambda(a,b|x,y)$, from which he derived his famous inequality satisfied by all local correlations. Let us emphasize that a violation of Bell's inequality not only tells us something about quantum physics, but - more impressively - tells us that in all possible future theories some spatially separated systems exhibit nonlocal correlations. Consequently, it is Nature herself that is nonlocal.

Many physicists feel uneasy with nonlocality\cite{nonrealism}. A part of the uneasiness comes from a confusion between nonlocal correlations and nonlocal signalling. The latter means the possibility to signal at arbitrarily fast speeds, a clear contradiction to relativity. However, the nonlocal correlations of quantum physics are nonsignalling. This should remove some of the uneasiness. Furthermore, note that in a nonsignalling world, correlations can be nonlocal only if the measurement results were not pre-determined. Indeed, if the results were pre-determined (and accessible with future theories and technologies), then one could exploit nonlocal correlations to signal. This fact has recently been used to produce bit strings with proven randomness \cite{StefanoRandom}. This is fascinating because it places quantum nonlocality no longer at the center of a debate full of susceptibilities and prejudice, but as a resource for future quantum technologies. We'll come back to this, but beforehand let us present a few recent experimental tests of quantum nonlocality.

The pioneering experiment by Clauser\cite{Clauser} suffered from a few loopholes, but these have since been separately closed\cite{locality, DetLoophole}. Still, correlations cry out for explanations, as emphasized by Bell\cite{BellCorrCry}. Everyone confronted with nonlocal correlations feels that the two systems somehow influence each other (e.g. Einstein's famous {\it spooky action at a distance}). This is also the way textbooks describe the process: a first measurement triggers a collapse of the entire state vector, hence modifying the state at the distant side. In recent years these intuitions have been taken seriously, leading to new experimental tests. If there is an influence from Alice to Bob, this influence has to propagate faster than light, as existing experiments have already demonstrated violation of Bell's inequality between space-like separated regions\cite{FinishMeasrmt}. But a faster than light speed can only be defined with respect to a hypothetical universal privileged reference frame, as the one in which the cosmic background radiation is isotropic. The basic idea is that if correlations are due to some "hidden influence" that propagates at finite speed, then, if the two measurements are sufficiently well synchronized in the hypothetical privileged frame, the influence doesn't arrive on time and one shouldn't observe nonlocal correlations. Remains the problem that one doesn't know {\it a priori} the privileged frame in which one should synchronize the measurements. This difficulty was recently circumvented by taking advantage of the Earth's 24 hours rotation, setting thus stringent lower bounds on the speed of these hypothetical influences\cite{SalartNature}. Hence, nonlocal correlations happen without one system influencing the oter. In another set of experiments the two observers, Alice and Bob, were set in motion in opposite directions in such a way that each in its own inertial reference frame felt he performed his measurement first and could thus not be influenced by his partner\cite{SuarezScarani97,MovingObs}. Hence, quantum correlations happen without any time-ordering.

All of today's experimental evidence points to one conclusion: Nature is nonlocal. As for any truly deep finding, this one has implications both for our world view and for future technologies. Let us first give an example of the second implication. Quantum Key Distribution (QKD) is the most advanced application of quantum information science. Today's commercial QKD systems rely on sound principles, but their implementation has to be thoroughly tested in order to check for unwanted side channels that an adversary could exploit. For example, the photons emitted by Alice could, in addition to carrying a quantum bit encoded in its polarization state, also carry redundant information unwittingly encoded in the timing of the photons, or in their spectra. This is possible because today's QKD systems do not rely on nonlocal correlations. If they would, the mere fact that the correlations between the data collected by Alice and Bob violate Bell's inequality would suffice to guarantee the absence of any side channel. This was the intuition of Ekert in 1991 \cite{Ekert91}, but was proven only in 2007 \cite{QKDDI}. Note the amazing consequence\cite{EkertPW}: in future, it will be possible to buy cryptography systems from ones adversary as the observation of nonlocal correlations will guarantee the proper functioning of the system!

To conclude let us come to the conceptual implications. In modern quantum physics entanglement is fundamental; furthermore, space is irrelevant - at least in quantum information science space plays no central role and time is a mere discrete clock parameter. In relativity space-time is fundamental and there is no place for nonlocal correlations. To put the tension in other words: no story in space-time can tell us how nonlocal correlations happen, hence nonlocal quantum correlations seem to emerge, somehow, from outside space-time.



\end{document}